\documentclass{PoS}

\PoS{PoS(LAT2005)042}

\usepackage{graphicx}
\usepackage[figuresright]{rotating}
\usepackage{color}
\let\normalcolor\relax

\title{Matrix elements and baryon spectroscopy from unquenched lattice QCD with improved staggered quarks.}

\ShortTitle{Baryon spectroscopy with improved staggered quarks.}

\author{
E.~B.~Gregory, A.~Irving, 
\speaker{C.~McNeile}, S.~Miller, 
and Z.~Sroczynski.
\\
Department of Mathematical Sciences, \\
University of\ Liverpool,\\ Liverpool\\  L69 3BX, UK\\
E-mail: \email{mcneile@amtp.liv.ac.uk}}

\abstract{
We look for the effect of open
decay channels on the masses of 
baryons in unquenched lattice QCD.
We apply variational smearing using fuzzed basis
states to a staggered nucleon operator.
The signal for
$\bar{s} s$ current in the nucleon is studied. 
The lattice calculations
are done using improved staggered fermions.
}

\FullConference{XXIIIrd International Symposium on Lattice Field Theory\\
                 25-30 July 2005\\
                 Trinity College, Dublin, Ireland}

\begin{document}

\section{Introduction}  \label{se:section}

There are a number of issues and questions about baryons
that can be addressed using lattice QCD. For example
the level ordering between the mass of the parity partner of the 
nucleon and the mass of the first excited state of the nucleon.
Also as we discuss in section~\ref{se:Content} there
are matrix elements of the nucleon, such as the scalar
current of strange quarks in the nucleon that have an uncertainty 
of at least a factor of 3. In this paper we make a start to
study the mass of the parity partner of the nucleon
and 
develop lattice techniques to 
compute the strangeness content of the nucleon
using staggered fermions.

Seduced by the light sea quark masses and large 
volumes of the unquenched lattice QCD calculations
performed by the MILC collaboration~\cite{Bernard:2001av,Aubin:2004wf}, 
we use the improved
staggered fermion formulation (known as ``Asqtad'') to do
our calculations. 
The unquenched calculations use
$2+1$ flavours of sea quarks with quark masses of
$0.007/0.05$. The lattice volume was $20^3\;64$ and
$\beta = 6.76$. We used the gauge configurations
generated by the MILC collaboration stored on the
public NERSC archive.

\section{The effect of open decays on the lattice.}

Many of the 
really interesting questions in hadronic
physics, such as the existence of hybrid mesons or glueballs, 
requires lattice QCD to be able to cope with hadrons
that decay via the strong interaction. Recent unquenched 
lattice QCD calculations now have light enough sea quarks
that hadrons are starting to decay.
The problem is most severe for particles that decay
via S-wave decays, because particles that decay via P-wave 
channels have additional kinematic obstacles because the 
decay products have to have momentum and
this can be quite large on typical lattices~\cite{MichaelCOLOR}. 

Indeed the MILC collaboration have claimed to see
problems for the $a_0$ state~\cite{Bernard:2001av,Aubin:2004wf} 
and $1^{-+}$ 
hybrid meson~\cite{Bernard:2003jd},
due to the quark masses being light enough for these
particles to decay into two mesons.
There have also been problems reported
with the singlet $0^{++}$ mesons~\cite{IrvingLAT05}. 
The experimental status of the above mesons
is confusing, so it would be better to study
the effect of hadronic decays on hadrons where
there is less uncertainty about the experimental data,
So it is important to study the issue of open decay
channels of baryons.

The masses of baryons quoted in the
particle data table (PDG)~\cite{Eidelman:2004wy}
are usually extracted using a partial waves analysis,
hence the angular momentum of 
the decay can be read off from the quantum 
numbers. In the table~\ref{tb:TTTTTT} we summarise some
data about some baryons from the 
PDG~\cite{Eidelman:2004wy}.
\begin{table}
\begin{tabular}{|c|c|c|}
\hline 
Baryon & Numbers & decay channel \\ \hline
N(1535)  &  $S_{11}$ & $N\pi$ , $N\eta$ \\
N(1440)  & $P_{11}$ &  $N\pi$,... \\
$\Lambda(1405)$  & $S_{01}$ &  $\Sigma\pi$,... \\
$\Delta$(1232) & $P_{33}$  & $N\pi$ \\
\hline
\end{tabular}
\label{tb:TTTTTT}
\end{table}
From table~\ref{tb:TTTTTT} the decay of the parity partner
of the nucleon (N(1535)) looks like a good candidate to have
the open decay of $N \pi$, in the current generation of unquenched
calculations.

There has been a lot of work on using quenched QCD 
to study
excited baryons. It has been found that 
large volumes ($> $ 3 fm ) are required,
as well as light quark masses for the chiral
extrapolation~\cite{Sasaki:2005ug}.
At the moment only the improved staggered program of the 
MILC collaboration comes close to these requirements.
Although in this calculation we have used a lattice with a 
physical box side of 2.4 fm, the MILC collaboration~\cite{Aubin:2004wf}
have also generated
unquenched gauge configurations with a box size of 3.4 fm.

Unfortunately, the identification of baryon
operators with physical states is non-trivial
in the 
staggered formalism~\cite{Golterman:1984dn}.
The
staggered baryon operators that are local in time
break down into irreducible representations
like:  $5 . \textcolor{red}{8} + 2 .\textcolor{red}{8^\prime} + 4 .\textcolor{red}{16}$~\cite{Golterman:1984dn}.
The nucleon couples to \textcolor{red}{8} and 
\textcolor{red}{16} representations~\cite{Golterman:1984dn}. 
Things are more complicated for the parity 
partner states. The $\Lambda(1405) \frac{1}{2}^{-}$ will
couple to the odd $\textcolor{red}{8}$ representation.
There are no interpolating operators for staggered
fermions that couple to the parity partner of the 
nucleon in the ground state.

The staggered parity partner to the nucleon with 
mass $m_{-}$ is extracted from the correlator $c(t)$ 
using the fit model in equation~\ref{eq:partChannel}.
\begin{equation}
c(t) = a_{+} \exp( -m_{+} t ) + \textcolor{red}{(-1)^t} a_{-} \exp(  -m_{-} t) 
\label{eq:partChannel}
\end{equation}
where $m_{+}$  is the mass of the nucleon.
There have been earlier attempts to extract the 
physical $m_{-}$ with unimproved staggered 
fermions~\cite{Ishizuka:1993mt}.

\begin{figure}[t]
\begin{center}
\leavevmode
\includegraphics[scale=0.3,clip,angle=270]{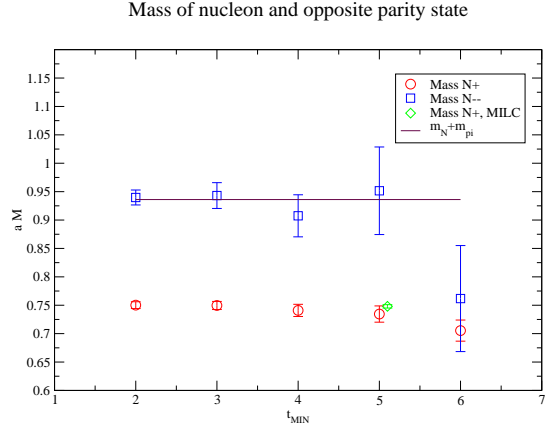}
\end{center}
\caption[]{\label{eq:paritypartner} {
Mass of the nucleon and the opposite parity state
that couples to the same correlator as a function of 
the first timeslice in the fit.~\ref{eq:partChannel}
}}
\end{figure}

In figure~\ref{eq:paritypartner} we plot the masses 
$m_{+}$ and $m_{-}$  in lattice units as a function
of the starting time in the fit. We used the 
gauge configurations from the MILC collaboration mentioned in the 
introduction and used wall sources in
Coulomb gauge to generate the quark propagators.
The actual number of configurations is in 
table~\ref{tb:FITresults}.
We also include the published number for 
$m_{N+}$ mass from MILC (they don't publish numbers
for $m_{N-}$) as well as the sum of the nucleon mass and
the mass of the Goldstone pion.

The data in the figure~\ref{eq:paritypartner} are
consistent with $m_{N-} = m_{N+} + m_{\pi}$. 
However, the box side has a physical length 
of 2.4 fm, so the $m_{N-}$ mass may be suppressed
by finite volume effects. In physical units the 
mass of the Goldstone pion is 310 MeV in this 
calculation. The physical
mass splitting between the nucleon and 
N(1535) state is about 600 MeV, so the state with the 
opposite parity in the nucleon channel is allowed
energetically (at least) to decay to a nucleon and pion.
A careful study of the $m_{N-}$ mass in larger
volumes and with heavier quark masses (where the decay is closed)
is required to confirm that this state has decayed.

We are not very comfortable with using equation~\ref{eq:partChannel}
to extract physical numbers for $m_{N-}$, because 
multiple exponential fits to a single channel can be
unstable. Also to study the issue of particle decay we need
as many excited states as possible. So we have started
to investigate using
a basis of
interpolating operators ($\Phi_i(t)$) and extracting
the masses and amplitudes via a variational technique.

A matrix of correlators is measured 
\begin{equation}
c(t)_{i \; j} = \langle \Phi_i(t) \Phi_j(0)^\dagger \rangle
\end{equation}
and analysed with a fit model of the form
\begin{equation}
c(t) = A^\dagger T A
\label{eq:FITmodel}
\end{equation}
For staggered fermions, the diagonal matrix $T$ contains exponentials
($e^{-m_{+} t}$)
and exponentials with prefactors of $(-1)^t$.
We prefer to fit the data to the expression in equation~\ref{eq:FITmodel},
rather than use techniques such as the  generalised 
eigenvalue method. 

We used ``fuzzed'' and local basis states~\cite{Lacock:1994qx}
originally developed for clover
fermions. To respect the staggered symmetries fuzzed
gauge links only separate the quark and anti-quark by multiples
of 2.
Here we try it for the local staggered nucleon
operator that belongs to the \textcolor{red}{8} representation.
\begin{equation}
O(t) = \sum_{x \; even} 
\epsilon^{abc}
\chi_a(x,t) \chi_b(x,t) \chi_c(x,t)
\end{equation}
The Chroma software system was used for the 
lattice calculations~\cite{Edwards:2004sx}.

In table~\ref{tb:FITresults} we show some results for
the variational analysis versus using wall sources
from MILC's analysis and our analysis. The variational
analysis is inconsistent at the 4$\sigma$ level with 
the results that use single wall sources. For Wilson fermions,
wall sources have small statistical errors, but tend
to approach a plateau very slowly. For this data set
MILC fitted the nucleon operator at smaller times
than for the equivalent analysis at other $\beta$ values.
It is not clear what causes the difference between the mass
from the wall source and variational analysis.

\begin{table}
\begin{tabular}{|c|c|c|c|c|}
\hline
Smearing & No. & Region & a $m_{+}$ & a $m_{-}$ \\
\hline
Wall     & $2 \times 342 $    & 4 - 15 & 0.74(1) & 0.91(4) \\
Variational     & $1 \times 230$    & 4 - 15 &  0.84(2) & 1.1(5) \\
Wall, MILC & $8 \times 493$    & 5 - 14 & 0.7480(30) &  \\
\hline
\end{tabular}
\caption{Fit results for the nucleon and the state with opposite 
parity in the same channel. The number before the 
$\times$ symbol is the number of time sources for the quark propagator
calculations on each
configuration}
\label{tb:FITresults}
\end{table}

\section{Strangeness content the nucleon} \label{se:Content}

The matrix element
$\langle N \mid  \overline{s} s \mid N \rangle $
is colloquially known as the 
strangeness content of the nucleon.
This matrix element 
is important for baryon chiral perturbation theory.
This matrix element is also used
in the 
neutralino-nucleon scalar cross-section~\cite{Bottino:1999ei} 
and hence crucial for the direct detection of dark matter.
It is convenient to discuss the ratios of matrix elements
\begin{equation}
y = 2 \frac{ \langle N \mid  \overline{s} s \mid N \rangle  }
         {  \langle N \mid  \overline{u}u + \overline{d}d  \mid N \rangle  }
\label{eq:ydefn}
\end{equation}
where $N$ is the nucleon operator.

In the table~\ref{tb:yREVIEW} we 
record some recent values for 
$y$ from lattice and non-lattice methods. 
There has also been a large quenched study 
of matrix elements related to $y$ by Lewis
et al.~\cite{Lewis:2002ix}.
The value of 
$y$ can range from 0.2 to 0.6.
The 
$\langle N \mid  \overline{s} s \mid N \rangle$  matrix 
element should not be 
effected by open decay channels, so a precision value
should be attainable.

\begin{table}
\begin{tabular}{|c|c||c|c|c|}
\hline 
Group & $y$  & Group &  $n_f$  &  $y$ \\ \hline
Borasoy and Mei{\ss}ner~\cite{Borasoy:1996bx} &   $0.21 \pm 0.20$ &
Kentucky~\cite{Dong:1995ec}  & 0 &  $0.36 (3)$ \\
Borasoy~\cite{Borasoy:1998uu} & $0.20 \pm 0.12 $ &
SESAM~\cite{Gusken:1998wy} & 2  & $0.59(13)$    \\ 
John Ellis~\cite{Ellis:2004cy} & $\sim 0.6$ &
UKQCD~\cite{Michael:2001bv} & 2 & $-0.28(33)$ \\ 
\hline
\end{tabular}
\caption{Some lattice and non-lattice determinations of $y$.}
\label{tb:yREVIEW}
\end{table}

On the lattice the matrix element $\langle N \mid \overline{s} s \mid
N \rangle $ is extracted from the correlation of a strange loop with
the two point nucleon correlator~\cite{Gusken:1998wy,Michael:2001bv}.
The loops were computed via a standard stochastic technique using
random noise sources~\cite{GregoryLAT05}. With staggered fermions the
scalar loop has no phase factors~\cite{IrvingLAT05}. 
However, the nucleon two point correlator does have a contribution
that oscillates in time (see equation~\ref{eq:partChannel}). 
We used the unquenched data set described in 
section~\ref{se:section}
with 330 gauge configurations. 
Quark propagators inverted from wall sources
were used.

The strange loop is inserted at time $t_1$.
The nucleon interpolating operators are at times $0$ and
$t$ with $t=t_1 + t_2$. The fit model for the
three point correlator is equation~\ref{eq:threePTdefn}.
\begin{equation}
   C_3^{(ab)}(t_1,t_2)=\sum_{i,j} c^{(a)}_i e^{-M_i t_1} x_{ij}
                  e^{-M_j t_2} c^{(b)}_j
\label{eq:threePTdefn}
\end{equation}
where $a$ and $b$ label the interpolating operators
and the time oscillating term is suppressed.
The amplitudes $c^{(b)}_j$ and masses $M_i$ are fixed
from the fits to the two point function.
The $\langle N \mid  \overline{s} s \mid N \rangle $ matrix
element is proportional to $x_{00}$.
In these preliminary fits we assumed $x_{01}$ =0, but
varied $x_{00}$ and $x_{11}$.
Using $t_1$ = 4 and fitting from t=9, we get
$x_{00} = 1.5 \pm 1.0 $. For this data set, the two point 
nucleon correlator
gets very noisy for $t > 12$. Given the large statistical
errors we do not quote a physical number for 
$\langle N \mid  \overline{s} s \mid N \rangle$ or
$y$.

\section{Conclusions}

We have started to study baryons on the 
lattice when the decay channels are open. 
It is critical to include finite volume studies
of these states~\cite{Sasaki:2005ug}. This is computationally feasible
with the current generation of improved staggered
fermion calculations.

For poorly known quantities such as
the strangeness content of the nucleon, we should be able to
(finally) pin this matrix element down. 
The statistical error on the bare matrix element is disappointingly
large, but we hope that by incorporating the variational
analysis and using many more gauge configurations the
errors should be reduced.


\begin{thebibliography}{10}

\bibitem{Bernard:2001av}
C.~W. Bernard {\em et.~al.}, {\it The qcd spectrum with three quark flavors},
  {\em Phys. Rev.} {\bf D64} (2001) 054506,
  [\href{http://xxx.lanl.gov/abs/hep-lat/0104002}{{\tt hep-lat/0104002}}].

\bibitem{Aubin:2004wf}
C.~Aubin {\em et.~al.}, {\it Light hadrons with improved staggered quarks:
  Approaching the continuum limit},  {\em Phys. Rev.} {\bf D70} (2004) 094505,
  [\href{http://xxx.lanl.gov/abs/hep-lat/0402030}{{\tt hep-lat/0402030}}].

\bibitem{MichaelCOLOR}
C.~Michael, {\it Hadronic decays},  {\em
  \href{http://pos.sissa.it/archive/conferences/020/008/LAT2005_008.pdf}{PoS(L%
AT2005)008}} (2005) [\href{http://xxx.lanl.gov/abs/hep-lat/0509023}{{\tt
  hep-lat/0509023}}].

\bibitem{Bernard:2003jd}
C.~Bernard {\em et.~al.}, {\it Lattice calculation of 1-+ hybrid mesons with
  improved kogut-susskind fermions},  {\em Phys. Rev.} {\bf D68} (2003) 074505,
  [\href{http://xxx.lanl.gov/abs/hep-lat/0301024}{{\tt hep-lat/0301024}}].

\bibitem{IrvingLAT05}
E.~Gregory, A.~Irving, C.McNeile, S.Miller, and Z.Sroczynski, {\it Scalar
  glueball and meson spectroscopy in unquenched lattice qcd with improved
  staggered quarks},  {\em
  \href{http://pos.sissa.it/archive/conferences/020/008/LAT2005_027.pdf}{PoS(L%
AT2005)027}} (2005).

\bibitem{Eidelman:2004wy}
{\bf Particle Data Group} Collaboration, S.~Eidelman {\em et.~al.}, {\it Review
  of particle physics},  {\em Phys. Lett.} {\bf B592} (2004) 1.

\bibitem{Sasaki:2005ug}
K.~Sasaki and S.~Sasaki, {\it Excited baryon spectroscopy from lattice qcd:
  Finite size effect and hyperfine mass splitting},  {\em Phys. Rev.} {\bf D72}
  (2005) 034502, [\href{http://xxx.lanl.gov/abs/hep-lat/0503026}{{\tt
  hep-lat/0503026}}].

\bibitem{Golterman:1984dn}
M.~F.~L. Golterman and J.~Smit, {\it Lattice baryons with staggered fermions},
  {\em Nucl. Phys.} {\bf B255} (1985) 328.

\bibitem{Ishizuka:1993mt}
N.~Ishizuka, M.~Fukugita, H.~Mino, M.~Okawa, and A.~Ukawa, {\it Operator
  dependence of hadron masses for kogut-susskind quarks on the lattice},  {\em
  Nucl. Phys.} {\bf B411} (1994) 875--902.

\bibitem{Lacock:1994qx}
{\bf UKQCD} Collaboration, P.~Lacock, A.~McKerrell, C.~Michael, I.~M. Stopher,
  and P.~W. Stephenson, {\it Efficient hadronic operators in lattice gauge
  theory},  {\em Phys. Rev.} {\bf D51} (1995) 6403--6410,
  [\href{http://xxx.lanl.gov/abs/hep-lat/9412079}{{\tt hep-lat/9412079}}].

\bibitem{Edwards:2004sx}
{\bf SciDAC} Collaboration, R.~G. Edwards and B.~Joo, {\it The chroma software
  system for lattice qcd},  \href{http://xxx.lanl.gov/abs/hep-lat/0409003}{{\tt
  hep-lat/0409003}}.

\bibitem{Bottino:1999ei}
A.~Bottino, F.~Donato, N.~Fornengo, and S.~Scopel, {\it Implications for relic
  neutralinos of the theoretical uncertainties in the neutralino nucleon
  cross-section},  {\em Astropart. Phys.} {\bf 13} (2000) 215--225,
  [\href{http://xxx.lanl.gov/abs/hep-ph/9909228}{{\tt hep-ph/9909228}}].

\bibitem{Lewis:2002ix}
R.~Lewis, W.~Wilcox, and R.~M. Woloshyn, {\it The nucleon's strange
  electromagnetic and scalar matrix elements},  {\em Phys. Rev.} {\bf D67}
  (2003) 013003, [\href{http://xxx.lanl.gov/abs/hep-ph/0210064}{{\tt
  hep-ph/0210064}}].

\bibitem{Borasoy:1996bx}
B.~Borasoy and U.-G. Meissner, {\it Chiral expansion of baryon masses and
  sigma-terms},  {\em Annals Phys.} {\bf 254} (1997) 192--232,
  [\href{http://xxx.lanl.gov/abs/hep-ph/9607432}{{\tt hep-ph/9607432}}].

\bibitem{Dong:1995ec}
S.~J. Dong, J.~F. Lagae, and K.~F. Liu, {\it $\pi n \sigma$ term , $\bar{s}s$
  in nucleon, and scalar form factor --- a lattice study},  {\em Phys. Rev.}
  {\bf D54} (1996) 5496--5500,
  [\href{http://xxx.lanl.gov/abs/hep-ph/9602259}{{\tt hep-ph/9602259}}].

\bibitem{Borasoy:1998uu}
B.~Borasoy, {\it Sigma-terms in heavy baryon chiral perturbation theory
  revisited},  {\em Eur. Phys. J.} {\bf C8} (1999) 121--130,
  [\href{http://xxx.lanl.gov/abs/hep-ph/9807453}{{\tt hep-ph/9807453}}].

\bibitem{Gusken:1998wy}
{\bf TXL} Collaboration, S.~Gusken {\em et.~al.}, {\it The pion nucleon
  sigma-term with dynamical wilson fermions},  {\em Phys. Rev.} {\bf D59}
  (1999) 054504, [\href{http://xxx.lanl.gov/abs/hep-lat/9809066}{{\tt
  hep-lat/9809066}}].

\bibitem{Ellis:2004cy}
J.~R. Ellis, {\it Today's view on strangeness},  {\em Eur. Phys. J.} {\bf
  A24S2} (2005) 3--10, [\href{http://xxx.lanl.gov/abs/hep-ph/0411369}{{\tt
  hep-ph/0411369}}].

\bibitem{Michael:2001bv}
{\bf UKQCD} Collaboration, C.~Michael, C.~McNeile, and D.~Hepburn, {\it The
  strangeness content of the nucleon},  {\em Nucl. Phys. Proc. Suppl.} {\bf
  106} (2002) 293--295, [\href{http://xxx.lanl.gov/abs/hep-lat/0109028}{{\tt
  hep-lat/0109028}}].

\bibitem{GregoryLAT05}
E.~B. Gregory, A.~C. Irving, C.~McNeile, S.~Miller, and Z.~Sroczynski, {\it
  Pseudoscalar singlet physics with staggered fermions},  {\em
  \href{http://pos.sissa.it/archive/conferences/020/008/LAT2005_083.pdf}{PoS(L%
AT2005)083}} (2005) [\href{http://xxx.lanl.gov/abs/hep-lat/0509193}{{\tt
  hep-lat/0509193}}].

\end{thebibliography}

\providecommand{\href}[2]{#2}\begingroup\raggedright\endgroup

\end{document}